\documentclass[aps,prl,preprint,showpacs,groupedaddress]{revtex4}
\usepackage {amssymb}
\usepackage {amsmath}
\usepackage {graphicx}
\usepackage {longtable}

\begin{document}
\title{Charge gap and ferroelastic lattice distortion in high-temperature
superconductors}

\author{Fedor V.Prigara}
\affiliation{Institute of Physics and Technology,
Russian Academy of Sciences,\\
21 Universitetskaya, Yaroslavl 150007, Russia}
\email{fvprigara@rambler.ru}

\date{\today}

\begin{abstract}

It is shown that a ferroelastic lattice distortion is associated
with the superconducting transition both in low- and
high-temperature superconductors. A low-temperature ferroelastic
transition in crystalline solids produces also a maximum in the
temperature dependence of their thermal conductivity. In layered
tetragonal superconductors with a sufficiently high transition
temperature, a ferroelastic distortion is improper and is caused
by an out-of-plane charge ordering. The relation between the
out-of-plane charge gap and the maximum superconducting transition
temperature for these superconductors is obtained.

\end{abstract}


\maketitle

A ferroelastic transition in a crystalline solid produces a spontaneous
temperature-dependent lattice distortion below the transition temperature
$T_{f} $ [1]. A ferroelastic lattice distortion is normally improper and is
associated with phase transitions of various types in crystalline solids
such as ferroelectric and antiferroelectric transitions [2], ferromagnetic
(magnetostriction) and antiferromagnetic transitions [2,3], and
metal-insulator transitions. In the last case, a ferroelastic lattice
distortion is a secondary effect of charge ordering (e.g., metal-insulator
transitions in $BiFeO_{3} $ [4], $Fe_{3} O_{4} $ [5], $BaVS_{3} $ [6],
$Ca_{2} RuO_{4} $ [7], $LiRh_{2} O_{4} $ [8], $LaMnO_{3} $ [9]). Here we
show that the superconducting transition both in low- and high-temperature
superconductors is also associated with a ferroelastic lattice distortion.
In anisotropic (layered) high-temperature superconductors (cuprate [10] and
pnictide [11] superconductors) with a sufficiently high transition
temperature $T_{c} $, this ferroelastic lattice distortion is improper and
is due to an out-of-plane charge ordering. The relation between the
out-of-plane charge gap $\Delta _{ch} $ and the maximum superconducting
transition temperature $T_{c} $ for cuprate and pnictide superconductors can
be inferred from a general criterion of phase transitions in crystalline
solids based on a concept of the critical number density of elementary
excitations [12]. We show also that there is a proper low-temperature
ferroelastic transition in crystalline solids which produces a maximum in
the temperature dependence of their thermal conductivity. Low-temperature
ferroelastic transitions in crystalline solids are consequences of a general
trend of lowering crystal symmetry with decreasing temperature [13].

Recently, it was shown [12,14] that, for phase transitions in crystalline
solids, there is a general relation between the energy $E_{0} $ of an
elementary excitation and the transition temperature $T_{c} $ of the form

\begin{equation}
\label{eq1}
E_{0} = \alpha T_{c} ,
\end{equation}

\noindent
where $\alpha \approx 18$ is a quantum constant. Here the Boltzmann constant
$k_{B} $ is included in the definition of the temperature \textit{T}.

The energy $E_{f} $ of an elementary ferroelastic excitation for a proper
ferroelastic transition is equal to the energy $\hbar \omega $ of the
optical phonon corresponding to the ferroelastic lattice distortion. The
last energy has an order of the Debye temperature $\Theta _{D} $, so that

\begin{equation}
\label{eq2}
E_{f} \cong \Theta _{D} .
\end{equation}

Now the equation (\ref{eq1}) gives the ferroelastic transition temperature $T_{f} $
for a proper ferroelastic transition in the form

\begin{equation}
\label{eq3}
T_{f} \cong \Theta _{D} /\alpha .
\end{equation}

There is a low-temperature maximum in the temperature dependence of the
thermal conductivity \textit{k(T)} for crystalline solids, both metals and
insulators [2]. The location of this maximum $T_{m} $ is close to the
transition temperature $T_{f} $ of the low-temperature ferroelastic
transition as given by the equation (\ref{eq3}), so that

\begin{equation}
\label{eq4}
T_{m} \approx T_{f} \cong \Theta _{D} /\alpha .
\end{equation}

For example, in diamond, the Debye temperature is $\Theta _{D} =
1860 K$ and the temperature of the thermal conductivity maximum is
$T_{m} \approx 100 K$ [2], in accordance with the relation
(\ref{eq4}). In copper (\textit{Cu}), the Debye temperature is
$\Theta _{D} = 310 K$ and the maximum of the thermal conductivity
is located at $T_{m} \approx 17 K$ [2]. In gallium arsenide
(\textit{GaAs}), the Debye temperature is $\Theta _{D} = 355 K$
and the maximum of the thermal conductivity occurs at $T_{m}
\approx 20 K$ [15].

The lattice thermal conductivity \textit{k} of a crystalline solid is
proportional to the heat capacity $c_{V} $ per unit volume, the mean speed
of sound $v_{s} $, and the mean free path $l_{ph} $of a phonon, as given by
the equation [2]

\begin{equation}
\label{eq5}
k = \left( {1/3} \right)c_{V} v_{s} l_{ph} .
\end{equation}

Below the temperature of the thermal conductivity maximum $T_{m} \approx
T_{f} $, the scattering of phonons is produced by ferroelastic domain
boundaries, so that the mean free path of a phonon is approximately
constant, $l_{ph} \cong const$, and the temperature dependence of the
thermal conductivity \textit{k(T)} is determined by the temperature
dependence of the heat capacity $c_{V} $,

\begin{equation}
\label{eq6}
c_{V} = \left( {12/5} \right)\pi ^{4}nk_{B} \left( {T/\Theta _{D}}
\right)^{3},
\end{equation}

\noindent
where \textit{n} is the number density of atoms. Thus, $k\left( {T} \right)
\propto T^{3}$ in this temperature range, except for the region of very low
temperatures where the electron thermal conductivity dominates in the case
of metals.

In copper (\textit{Cu}), the mean speed of sound (in the Debye
model) is $v_{s} = 2.6 kms^{ - 1}$ and the maximum value of the
thermal conductivity $k_{m} = 50 W/cmK$ (at $T_{m} \approx \Theta
_{D} /\alpha $) gives an estimation of the mean free path of a
phonon below $T_{m} $ at the level of $l_{ph} \approx 0.1 mm$.
\textit{Cu} has an fcc crystal structure. The ferroelastic lattice
distortion in \textit{Cu} below the temperature of the thermal
conductivity maximum $T_{m} \approx T_{f} $ is presumably
rhombohedral (many metals, such as Co, Ti, Zr, rare earth metals,
have a low-temperature hcp phase and high-temperature fcc and bcc
phases). Note that, due to the Wiedemann-Franz law, the
low-temperature maximum in the thermal conductivity of metals
cannot be attributed to the electron thermal conductivity, since
there is no maximum in the electrical conductivity of metals at
these temperatures.

The temperature-dependent ferroelastic lattice distortion $\delta = \Delta
a/a$ (where \textit{a} is a lattice parameter), which is increasing with the
decreasing temperature, gives a negative contribution to the thermal
expansion coefficient $\alpha _{L} = \left( {1/L} \right)dL/dT$ (where
\textit{L} is a linear size of a crystal sample), similarly to
magnetostriction in ferromagnetic alloys, and can produce an overall
negative thermal expansion. Such is the case in gallium arsenide
(\textit{GaAs}) below 55 \textit{K} [15] and in silicon (\textit{Si}) at low
temperatures. Since the linear thermal expansion coefficient $\alpha _{L} $
has an order of $10^{ - 6}K^{ - 1}$ in \textit{GaAs} and \textit{Si} at low
temperatures, the maximum lattice distortion is rather small, $\delta < 10^{
- 4}$.

A negative thermal expansion is also observed in some superconductors
($MgB_{2} $, \textit{Ta}) below the superconducting transition temperature
$T_{c} $ [16] and is indicative of a temperature-dependent ferroelastic
lattice distortion associated with the superconducting phase transition.

There is a close relation between superconductivity and low-temperature
ferroelastic transitions, since the superconducting state is produced by the
interaction between the electrons and lattice excitations. In the case of a
proper ferroelastic lattice distortion associated with the superconducting
transition, the energy $E_{s} $ of an elementary superconducting excitation
is close to the energy $E_{f} $ of an elementary ferroelastic excitation, so
that

\begin{equation}
\label{eq7}
E_{s} = \alpha T_{c} \approx E_{f} \cong \Theta _{D} .
\end{equation}

The equation (\ref{eq7}) gives the relation between the maximum superconducting
transition temperature $T_{c} $ and the Debye temperature in the form

\begin{equation}
\label{eq8}
T_{c} \cong \Theta _{D} /\alpha .
\end{equation}

In cuprate and pnictide superconductors, the superconducting transition
temperature depends on the doping level and on deviations from the
stoichiometric composition. The equation (\ref{eq8}) corresponds in these cases to
the maximum transition temperature which is achieved at the optimal doping
level.

The relation (\ref{eq8}) is valid for many superconductors, both
low- and high-temperature ones, such as $MgB_{2} $ ($T_{c} = 39
K$, $\Theta _{D} = 800 \pm 80 K$) [12], $Pr_{2} Ba_{4} Cu_{7}
O_{15 - \delta}  $ ($T_{c} = 16 K$, $\Theta _{D} = 340 K$) [17],
\textit{Pb} ($T_{c} = 7.2 K$, $\Theta _{D} = 90 K$), \textit{Hg}
($T_{c} = 4.15 K$, $\Theta _{D} = 96 K$), and many other elemental
superconductors (see data on the maximum transition temperature
achievable in elemental superconductors under pressure in Ref.
18).

The relation (\ref{eq8}) seems to be valid in electron-doped
cuprate superconductors [17]. However, in hole-doped cuprate
superconductors [19] and in recently discovered pnictide
superconductors [11], the maximum transition temperature $T_{c} $
normally exceeds the value given by the equation (\ref{eq8}). For
example, in $Ba_{0.5} K_{0.5} Fe_{2} As_{2} $, the Debye
temperature is $\Theta _{D} = 246 K$, whereas the transition
temperature is $T_{c} = 38 K$ [11].

The properties of pnictide superconductors are similar to those of
previously known high- temperature superconductors. For example,
the temperature dependence of heat capacity \textit{c(T)} in
$Ba_{0.5} K_{0.5} Fe_{2} As_{2} $ [11] exhibits a broad maximum at
the temperature $T_{AFM}^{ *}  = \alpha _{P} T_{c} \approx 14 K$,
where $\alpha _{P} \approx 3/8$ is the atomic relaxation constant
[12]. This maximum in the temperature dependence of the heat
capacity is caused by the contribution from antiferromagnetic
fluctuations in the superconducting phase [12]. There is a similar
maximum in the \textit{c(T)} dependence below $T_{c} $ in $MgB_{2}
$ [16].

The value of the atomic relaxation constant $\alpha _{P} \approx
3/8$ is characteristic for high-temperature superconductors and
can be attributed to their anisotropic layered crystal structure.
In isotropic low-temperature superconductors, the value of the
atomic relaxation constant is $\alpha _{P} \approx 3/16$ [12]. The
in-plane penetration depth $\lambda _{ab} $ in pnictide
superconductors [20] has an order of the size of a crystalline
domain $d_{c} $, $\lambda _{ab} \cong d_{c} \approx 180 nm$ [12],
as is the case in other high-temperature superconductors.

In layered high-temperature superconductors with $T_{c} > \Theta _{D}
/\alpha $, a ferroelastic lattice distortion is improper and is caused by an
out-of-plane charge ordering. In this case, the energy $E_{f} $ of an
elementary ferroelastic excitation is close to the magnitude of the
out-of-plane charge gap $\Delta _{ch} $, so that

\begin{equation}
\label{eq9}
E_{s} = \alpha T_{c} \cong E_{f} \cong \Delta _{ch} ,
\end{equation}

\noindent
and the maximum superconducting transition temperature $T_{c} $ is
determined by the relation

\begin{equation}
\label{eq10}
T_{c} \cong \Delta _{ch} /\alpha .
\end{equation}

The out-of-plane charge gap has been directly observed in the
scanning tunnelling microscopy measurements performed on single
crystals of a hole-doped cuprate superconductor $Bi_{2} Sr_{2}
CaCu_{2} O_{8 + \delta}  $ in the direction perpendicular to the
$CuO_{2} $ planes [10]. There is a modulation in the measured gap
due to the lattice modulation with a period of $2.6nm$along the
orthorhombic a-axis (atoms are displaced in the c-axis direction).
There is also a disorder in the gap maxima which occurs in
association with the non-stoichiometric dopant oxygen atom
locations, with gap values strongly increased in their vicinity.
If the maximum transition temperature is $T_{c} \cong 80 K \approx
7meV$ for these samples, then the energy $E_{s} $ of an elementary
superconducting excitation is $E_{s} = \alpha T_{c} \cong 0.125
eV$ and is close to the maximum measured mean charge gap $\Delta
_{ch} = 0.114 eV$. The last value does not include the disorder in
the charge gap maxima ranging up to $0.140 eV$.

The out-of-plane charge gap $\Delta _{ch} $ seems to be related to the
out-of-plane Coulomb gap studied theoretically in Ref. 21.

With the enhancement of the distance between the $CuO_{2} $ planes in
hole-doped cuprate superconductors, the out-of-plane charge gap $\Delta
_{ch} $ tends to increase, and so does the maximum transition temperature
$T_{c} $, in accordance with the relation (\ref{eq10}). The record transition
temperatures were achieved in multi-layered \textit{Hg}-based cuprate oxides
[18].

To summerize, we show that, due to the lowering of crystal symmetry with
decreasing temperature, a low-temperature ferroelastic transition in
crystalline solids occurs producing a maximum in the temperature dependence
of their thermal conductivity. Superconductivity is closely related to
low-temperature ferroelastic transitions in crystalline solids, the maximum
superconducting transition temperature being close to the corresponding
ferroelastic transition temperature. This ferroelastic transition is proper
in the case of low-temperature superconductors and some high-temperature
superconductors such as magnesium diboride. For most of high-temperature
superconductors, the corresponding ferroelastic transition is improper and
is caused by an out-of-plane charge ordering. There is a direct experimental
evidence for this out-of-plane charge gap in hole-doped cuprate
superconductors.

\begin{center}
---------------------------------------------------------------
\end{center}

[1] S.H.Curnoe and A.E.Jacobs, Phys. Rev. B \textbf{64}, 064101 (2001).

[2] G.S.Zhdanov, \textit{Solid State Physics} (Moscow University Press,
Moscow, 1961) [G.S.Zhdanov, \textit{Crystal Physics} (Oliver \& Boyd,
Edinburgh, 1965)].

[3] R.J.McQueeney, J.-Q.Yan, S.Chang, and J.Ma, Phys. Rev. B (accepted),
arXiv:0806.3576 (2008).

[4] R.Palai, R.S.Katiyar, H.Scmid et al., Phys. Rev. B \textbf{77}, 014110
(2008).

[5] D.Mertens, \textit{The Verwey Transition in Magnetite} (unpublished).

[6] I.Kezsmarki, G.Mihaly, R.Gaal, N.Barisic, H.Berger, L.Forro, C.C.Homes,
and L.Mihaly, Phys. Rev. B \textbf{71}, 193103 (2005).

[7] S.G.Ovchinnikov, Usp. Fiz. Nauk \textbf{173}, 27 (2003) [Physics-Uspekhi
\textbf{46}, 21 (2003)].

[8] Y.Okamoto, S.Niitaka, M.Uchida et al., Phys. Rev. Lett. \textbf{101},
086404 (2008).

[9] H.Zenia, G.A.Gehring, and W.M.Temmerman, New J. Phys. \textbf{7}, 257
(2005).

[10] J.A.Slezak, J.Lee, M.Wang et al., Proc. Nat. Acad. Sci. USA
\textbf{105}, 3203 (2008).

[11] J.K.Dong, L.Ding, H.Wang, X.F.Wang, T.Wu, X.H.Chen, and S.Y.Li,
arXiv:0806.3573 (2008).

[12] F.V.Prigara, Phys. Rev. Lett. (submitted), arXiv:0708.1230 (2007).

[13] V.S.Urusov, \textit{Theoretical Crystal Chemistry} (Moscow University
Press, Moscow, 1987).

[14] F.V.Prigara, Phys. Rev. Lett. (submitted), arXiv:0805.4325 (2008).

[15] N.G.Ryabtsev, \textit{Materials for Quantum Electronics} (Soviet Radio
Publishers, Moscow, 1972).

[16] J.J.Neumeier, T.Tomita, M.Debessai, J.S.Schilling, P.W.Barnes,
D.G.Hinks, and J.D.Jorgensen, Phys. Rev. B \textbf{72}, 220505 (2005).

[17] A.Matsushita, K.Fukuda, Y.Yamada, F.Ishikawa, S.Sekiya, M.Hedo, and
T.Naka, Sci. Technol. Adv. Mater. \textbf{8}, 477 (2007).

[18] J.S.Schilling, in \textit{Handbook of High Temperature
Superconductivity: Theory and Experiment}, editor J.R.Schrieffer, associated
editor J.S.Brooks (Springer Verlag, Hamburg, 2007).

[19] B.Chen, S.Mukhopadhyay, W.P.Halperin, P.Guptasarma, and D.G.Hinks,
Phys. Rev. B \textbf{77}, 052508 (2008).

[20] F.L.Pratt, P.J.Baker, S.J.Blundell et al., arXiv:0910.0973 (2008).

[21] J.Halbritter, Physica C \textbf{302}, 221 (1998).

\end{document}